\input{aipcheck}

\documentclass[
   ,final            
 ]
 {aipproc}

\layoutstyle{6x9}

\usepackage{amsfonts}%
\usepackage{mathrsfs}

\def\Barcelo{Barcel\'o}%
\def\e{{\mathrm e}}%
\def\g{{\mbox{\sl g}}}%
\def\d{{\mathrm d}}%
\def\ie{{\em i.e.\/}}%
\newcommand{\scri}{\mathscr{I}}

\begin{document}

\title{Revisiting the semiclassical gravity scenario 
for gravitational collapse}






\author{Carlos \Barcelo}{
 address={Speaker. Inst. Astrof\'{\i}sica de Andaluc\'{\i}a, CSIC,
Camino Bajo de Hu\'etor 50, 18008 Granada, Spain}
}

\author{Stefano Liberati}{
 address={Int. School for Advanced Studies, Via Beirut
2-4, 34014 Trieste, Italy and INFN, Sezione di Trieste}
}

\author{Sebastiano Sonego}{
 address={Universit\`a di Udine, Via delle Scienze 208, 33100 Udine, Italy}
}

\author{Matt Visser}{
 address={School of Maths, Statistics, and Operations Research, Victoria U. of Wellington,
New Zealand}
}

\begin{abstract}

The existence of extremely dark and compact astronomical bodies is by
now a well established observational fact. On the other hand,
classical General Relativity predicts the existence of black holes
which fit very well with the observations, but do lead to important
conceptual problems. In this contribution we ask ourselves the
straightforward question: {\em Are the dark and compact objects that
we have observational evidence for black holes in the sense of General
Relativity?} By revising the semiclassical scenario of stellar
collapse we find out that as the result of a collapse some alternative
objects could be formed which might supplant black holes.

\end{abstract}

\maketitle


\section{Introduction}

An already impressive quantity of high-quality observational data
points towards the existence of very dark and compact astronomical
bodies. On the theoretical side, classical General Relativity (GR)
predicts under quite general and reasonable conditions (some
positive-energy condition applied to the matter stress-energy-momentum
tensor) that stars of large masses (greater than about 3 solar masses)
cannot avoid complete collapse once they have exhausted their nuclear
fuel. When the collapsing star reaches a radius smaller than its
gravitational radius (\ie, its Schwarzschild radius $r_g=GM/c^2$ if it
is non-rotating) there is a region of spacetime that cuts itself off
forever from the rest of the universe; nothing, not even light can
escape from this region. A black hole in GR is defined as precisely
this cut-off region. The surface that separates this region from the
rest of the universe is called the black-hole event horizon.

The notion of black hole fits so nicely with the observations that the
extremely compact and dark objects found out there in the sky are
customarily called black holes or more gingerly black hole
candidates. But are they really black holes in the precise sense of
classical GR? At the same time that classical GR predicts the
existence of black holes it also predicts that inside the event
horizon of a standard (non-extremal) black hole there will always be a
singularity. The formation of singularities is a worrisome feature 
of classical GR that to many people suggests one has to unavoidably 
incorporate quantum ingredients into the study of gravity. 

One of the most conservative ways of incorporating quantum aspects
into GR is to treat the geometry as classical but the matter as
quantum, with the expectation value of the stress-energy-momentum
tensor (SET) in the specific quantum state of matter acting as the
source of gravity. This framework constitutes the so-called
semiclassical GR. It is standard to write the semiclassical Einstein
equations as 
\begin{eqnarray}%
G_{\mu\nu}=8\pi\left(T^c_{\mu\nu}+T^Q_{\mu\nu}\right)~, 
~~~~~~{\rm with}~~~~~~ 
T^Q_{\mu\nu} \equiv \langle \psi|\widehat T_{\mu\nu}| \psi \rangle~.  
\label{sceq}%
\end{eqnarray}%
Here, one separates the
expectation value of the total SET into a classical part,
$T^c_{\mu\nu}$, and a purely quantum contribution calculated through
renormalization. Significant deviations from the standard Einstein
equations can appear only if the renormalized SET in equation
(\ref{sceq}) becomes comparable with the classical SET. Ideally, in
this approach one should self-consistently solve the semiclassical
Einstein equations, but this is in general rather
complicated. Instead, what is typically done is to take a fixed
geometry, calculate the renormalized stress-energy-momentum tensor
(RSET) of the quantum fields in specific states and analyze what the
effects of this RSET, taken as an additional source of gravity, would
be.

Within semiclassical GR it was found that black holes cannot be
stationary~\cite{boulware} but that they must
evaporate~\cite{hawking-ter}. Conceptually, these evaporating black
holes might be very different from classical black holes, in the sense
that they might have no strict event horizon after
all~\cite{hawking-info}. However, the evaporation of a stellar-mass
black hole is so slow that in astrophysical terms the latter would be
completely indistinguishable from a classical black hole. Instead of a
strict event horizon it would possess a long-living trapping horizon.
A light signal might escape from the evaporating black hole, but only
after the evaporation process has developed sufficiently, which will
take many times the present age of the universe.

In this contribution we want to take one step back and ask ourselves
whether semiclassical GR leads unavoidably to the formation of
evaporating black holes or, in other words, whether it can provide
qan alternative route producing dark and compact stellar-mass objects
having, however, no horizons of any kind. These objects might in
principle be suitable for astrophysical exploration (either detection 
or elimination) in a reasonable time frame.

\section{The results}

Let us revisit the semiclassical GR scenario for stellar collapse.
Consider a star of mass $M$ in hydrostatic equilibrium in empty
space. For such a configuration the appropriate quantum state is well
known to be the Boulware vacuum state $|0_{\rm
B}\rangle$~\cite{boulware}, which is defined unambiguously as the
state with zero particle content for static observers. This quantum
state is regular everywhere both inside and outside the star (this
state is also known as the static, or Schwarzschild,
vacuum~\cite{birrell-davies}). If the star is sufficiently dilute (so
that its radius is very large compared to $2M$), then the spacetime is
nearly Minkowskian and such a state will be virtually
indistinguishable from the Minkowski vacuum. Hence, the expectation
value of the RSET will be
negligible throughout the entire spacetime. This is the reason why,
when calculating the spacetime geometry associated with a dilute star,
one only needs to care about the classical contribution to the SET.

Imagine now that, at some moment, the star begins to collapse. The
evolution proceeds as in classical general relativity, but with some
extra contributions as spacetime dynamics will also affect the
behaviour of any quantum fields that are present, giving place to
\emph{both} particle production \emph{and} additional vacuum polarization
effects.  Contingent upon the standard scenario being correct, if we
work in the Heisenberg picture there is a single globally defined
regular quantum state $|C\rangle =|\mathrm{collapse}\rangle$ that
describes these phenomena.

If the collapse proceeds very quickly, with the entire structure
almost freely falling into itself, then it has (numerically) been
shown (see for example~\cite{pp}) that the RSET maintains negligible
values throughout the collapse, well beyond the moment at which the
trapping horizon forms. One can neglect these quantum effects insofar
as the dynamics of the collapse is concerned, at least up to the
appearance of a trapped region. Only when the collapse generates large
curvatures, that is, near the incipient singularity inside the trapped
region, do quantum effects become important. The only residual quantum
effect that can in principle be observable from outside is the
presence of Hawking radiation coming out from the trapping horizon,
and the consequent shrinking of the trapped region. Remember
nonetheless that both these effects are almost unnoticeable in an
astrophysical context.

This constitutes the standard view of the collapse process in
semiclassical GR.  However, we found~\cite{fate-of-collapse}
that if for whatever reason the collapse deviates significantly
from free fall, then the RSET can contribute appreciably to
the dynamics of collapse itself, {\em before} any trapped region has
formed. In these situations the RSET provides an energy-condition
violating contribution that will tend to slow down further the
collapse. This opens up the possibility that vacuum polarization
effects, encoded in the RSET, delay indefinitely the appearance of
a trapped region by completely halting the collapse, or by producing an
asymptotic approach to trapping horizon formation. In the latter
case we also showed~\cite{quasi-particle-prl} that the resulting
object might maintain most of the thermodynamic properties of black
holes, as it could evaporate by emitting a Planckian spectrum of
particles with an associated temperature indistinguishable in practice 
from the Hawking temperature.

If this was finally the route taken by nature, the new quantum-corrected
bodies that substitute classical black holes would not be
evaporating black holes, as the standard view maintains, but ``black
stars'' --- objects radically different from black holes in several
respects. They would be material bodies, with a real and observable
(although extremely red-shifted) surface and a non-empty interior,
filled with matter at a density at least one order of magnitude greater
than that of a neutron star. Having no horizons, they would not put any
fundamental barrier to their complete astrophysical exploration. They
would be supported by the most basic form of quantum pressure, the one
provided by vacuum polarization itself. 

The existence of this huge vacuum polarization energy would be tied up
to the fact of having a configuration maintaining itself close to
horizon formation, and not to the specific values of the curvature in
this same region. In the case that they Hawking-like evaporate, (we
don't know whether or not this might require some fine-tuning), the
internal structure of these bodies would be that of a rainbow of
temperature: Imagining the body as composed of thin spherical shells,
each of these will be slowly shrinking, chasing but never reaching its
own horizon; in this way to each shell one associates a Hawking-like
temperature which steadily increases towards the centre.

\section{The calculation}

For simplicity, consider a massless quantum field and
restrict the analysis to spherically symmetric solutions.  Every
mode of the field can (neglecting back-scattering) be described as a
wave coming in from $\scri^-$ (\ie, from $r\to +\infty$, $t\to
-\infty$), going inwards through the star till bouncing at its
center ($r=0$), and then moving outwards to finally reach $\scri^+$.
As in this paper we are going to work in $1+1$ dimensions (\ie, we
shall ignore any angular dependence), for 
convenience instead of considering wave reflections at $r=0$ we will
take two mirror-symmetric copies of the spacetime of the collapsing
star glued together at $r=0$. In one copy
$r$ will run from $-\infty$ to $0$, and in the other from $0$ to
$+\infty$. Then one can concentrate on how the modes change on their
way from $\scri^-_\mathrm{left}$ (\ie, $r\to -\infty$, $t\to
-\infty$) to $\scri^+_\mathrm{right}$ (\ie, $r\to +\infty$, $t\to
+\infty$). Hereafter, we will always implicitly assume this
construction and will not explicitly specify ``left'' and ``right''
except  where it might cause confusion.

With reference to this construction, we shall start by considering a
set of affine coordinates $U$ and $W$, defined on
$\scri^-_\mathrm{left}$ and $\scri^-_\mathrm{right}$
respectively. These coordinates are globally defined over the
spacetime and the metric can be written as
\begin{equation}
\g =  -C(U,W)\, \d U \, \d W\;.
\end{equation}
Given that we shall be concerned with events which lie outside of
the collapsing star on the right-hand side of our diagram, we can
also choose a second double-null coordinate patch $(u,W)$, where $u$
is taken to be affine on $\scri^+_\mathrm{right}$, in terms of which
the metric is
\begin{equation}
\g =  -\bar C(u,W) \, \d u \, \d W~,
~~~~~~{\rm with}~~~~~~
C(U,W) = {\bar C(u,W)/ \dot p(u) }~.
\end{equation}
Of course $U=p(u)$ describes the coordinate transformation which
in turn is the red-shift function, and encodes how a wave packet 
of initial frequency $\omega'$ (on $\scri^-_\mathrm{left}$) gets red-shifted in passing in, through, and out
the collapsing star: on $\scri^+_\mathrm{right}$, $\omega(u, \omega')=\dot p(u) \omega'$.
Furthermore, as long as we are outside  the collapsing star it is
safe to assume that a Birkhoff-like result holds, and take $\bar
C(u,W)$ as being that of a static spacetime.

Now for \emph{any\/} massless quantum field, the RSET (corresponding
to a quantum state that is initially Boulware) has
components~\cite{birrell-davies, dfu}
\begin{equation}
T^Q_{UU} \propto C^{1/2}\, \partial_U^2\, C^{-1/2}~;~~~~~~~
T^Q_{WW} \propto C^{1/2}\, \partial_W^2\, C^{-1/2}~;~~~~~~~
T^Q_{UW} \propto R~.
\end{equation}
The coefficients arising here are not particularly important, and
will in any case depend on the specific type of quantum field under
consideration.

The components $T^Q_{WW}$ and $T^Q_{UW}$ will necessarily be well
behaved throughout the region of interest; in particular they are
the same as in a static spacetime and are known to be regular. On
the contrary $T^Q_{UU}$ shows a more complex structure due to the
non-trivial relation between $U$ and $u$. A brief computation yields
\begin{equation}
C^{1/2}\, \partial_U^2\, C^{-1/2} = {1\over\dot p^2} \left[
\bar C^{1/2}\, \partial_u^2\, \bar C^{-1/2} - \dot p^{1/2}\, \partial_u^2\, \dot p^{-1/2}
\right].
\label{T}
\end{equation}
The key point here is that we have two terms, one ($\bar C^{1/2}\,
\partial_u^2\, \bar C^{-1/2}$) arising purely from the static
spacetime outside the collapsing star, and the other ($\dot
p^{1/2}\, \partial_u^2\, \dot p^{-1/2} $) arising purely from the
dynamics of the collapse. If, and only if, the horizon is assumed to
form at finite time will the leading contributions of these two
terms cancel against each other --- this is the standard scenario.

Indeed the first term is exactly what one would compute from using
standard Boulware vacuum for a static star. As the surface of the
star recedes, more and more of the  static spacetime is
``uncovered'', and one begins to see regions of the spacetime where
the Boulware contribution to the RSET is more and more negative, in
fact diverging as the surface of the star crosses the horizon.

To study the regular or divergent behaviour of the RSET in different
geometries it is convenient to use a set of regular coordinates. 
Let us choose Painlev\'e-Gullstrand coordinates in which the metric 
is expressed as 
\begin{eqnarray}
ds^2 =-\left[c^2-v^2(t,x)\right]\d t^2
-2\,v(t,x)\,\d t\,\d x+\d x^2~. 
\end{eqnarray}
Taking the convention $v<0$, in these coordinates the trapping
horizon is located at the point at which $c+v=0$. Then, the components
of the RSET which can be potentially divergent are $T^Q_{tx}$ and 
$T^Q_{xx}$. Explicitly, the terms that can be divergent are
\begin{eqnarray}
T^Q_{tx} = -{\dot p^2\over c+v} \,T^Q_{UU} + \cdots
\label{tx}
\\
T^Q_{xx} = {\dot p^2\,\over (c+v)^2}\,T^Q_{UU} + \cdots
\label{xx}
\end{eqnarray}
%

\noindent
{\bf Calculation assuming normal horizon formation}
\vspace{0.2cm}

Hereafter, we shall for simplicity restrict our attention to the
case $c(x)\equiv 1$.  Placing the horizon at $x=0$ for convenience,
we can write the asymptotic expansion
\begin{equation}%
v(x) = -1 + \kappa\, x + \kappa_2\, x^2 + \cdots\;,%
\label{v}%
\end{equation}%
where $\kappa$ can be identified with the surface
gravity~\cite{quasi-particle-prl,analogue}.

Consider first the static Boulware term in equation (\ref{T}). We
have (placing the horizon at $x=0$ for convenience)
\begin{equation}
\bar C = -\frac{\dot{p}}{U_x\,W_x} = -{1\over u_x\, W_x} = 1 -
v(x)^2 \approx 2\, \kappa\, x\;.
\end{equation}
The relevant derivative in $\partial_u$ is then that with respect to
$x$, and we can write
\begin{eqnarray}
\bar C^{1/2}\, \partial_u^2\, \bar C^{-1/2} \approx
(2\,\kappa\, x)^{1/2}\,  \kappa\, x\, \partial_x \left( \kappa\, x\, \partial_x (2\,\kappa\, x)^{-1/2} \right)= \kappa^2/4\;.
\end{eqnarray}
In fact, keeping the sub-leading terms one finds
\begin{equation}
\bar C^{1/2}\, \partial_u^2\, \bar C^{-1/2} = {\kappa^2\over4} + \mathcal{O}(x^2).
\end{equation}
By equations (\ref{tx}) and (\ref{xx}), it is clear that because of
the constant term $\kappa^2/4$, the components $T_{tx}$ and $T_{xx}$
of the RSET contain contributions that diverge as $x^{-1}$ and
$x^{-2}$, respectively, as $x\to 0$. (The sub-leading terms lead to
finite contributions of order $\mathcal{O}(x)$ and $\mathcal{O}(1)$
respectively.)

In counterpoint, assuming horizon formation, let us now calculate
the dynamical contribution to the RSET ($\dot p^{1/2}\,
\partial_u^2\, \dot p^{-1/2} $). It is well known that any
configuration that produces a horizon at a finite time $t_{\rm H}$
leads to an asymptotic (large $u$) form
\begin{eqnarray}%
p(u)\approx U_{\rm H} - A_1\; \e^{-\kappa u}\;,
\label{standard-p}
\end{eqnarray}%
where $U_H$ and $A_1$ are suitable constants. Taking into account
the asymptotic expression (\ref{v}) for $v(x)$ near $x=0$, it is
very easy to see that the potential divergence at the horizon due to
the static term is exactly cancelled by the dynamical term. In this
way we have recovered the standard result that the RSET at the
horizon of a collapsing star is regular.

However, the previous relation is an asymptotic one, and for what we
are most interested in (the value of the RSET close to horizon
formation) it is important to take into account extra terms that
will be sub-dominant at late times. Indeed, we can describe the
location of the surface of a collapsing star that crosses the
horizon at time $t_{\rm H}$ by
\begin{equation}
x= r(t) -2M = \xi(t) = -\lambda (t-t_{\rm H})+\cdots\;,
\end{equation}
where the expansion makes sense for small values of $|t-t_{\rm H}|$,
and $\lambda$ represents the velocity with which the surface crosses
the gravitational radius.  Let $t_0$ be the time at which a
right-moving light ray corresponding to  null coordinates $u$ and
$U$ crosses the surface of the star.  Then on the one hand
\begin{eqnarray}%
t_f-t_0=\int_{\xi(t_0)}^{x_f} \frac{\d x'}{1+v(x')}\;,
\end{eqnarray}%
which for $t_0\approx t_{\rm H}$ (implying $r(t_0)\approx 2M$) leads to the approximate expression for $u:=\lim_{t_f\to +\infty}\left(t_f-x_f\right)$
\begin{eqnarray}%
u \approx \left(t_0-t_{\rm H}\right) - \frac{1}{\kappa}\ln \left(-\lambda \left(t_0-t_{\rm H}\right)\right) +C_1\;,
\end{eqnarray}%
so that
\begin{equation}
t_0 - t_{\rm H} \approx
C_2 \,{\e^{-\kappa u}\over\lambda} + \cdots
\label{t0-tH}
\end{equation}
On the other hand, since $U(t_0)$ is simply some regular function, we have
\begin{equation}
U(t_0) = U_{\rm H} + U'_{\rm H} \; (t_0-t_{\rm H}) + {U''_{\rm H}\over2} \; (t_0-t_{\rm H})^2 +\cdots
\label{Ut0}
\end{equation}
Inserting (\ref{t0-tH}) into (\ref{Ut0}) we obtain an asymptotic expansion
\begin{eqnarray}%
p(u) = U_{\rm H} - A_1 \; \e^{-\kappa u} + {A_2\over 2}\;  \e^{-2\kappa u}
+ {A_3\over 3!} \; \e^{-3\kappa u}+ \cdots~.
\label{expp}
\end{eqnarray}%
Then
\begin{eqnarray}
\dot p^{1/2}\, \partial_u^2\, \dot p^{-1/2}
=
\frac{\kappa^2}{4} + \left[ -{1\over2} {A_3\over A_1} + {3\over4} \left({A_2\over A_1}\right)^2 \right]\kappa^2\, \e^{-2\kappa u}
+ \vphantom{{1\over4}} \mathcal{O}\left( \e^{-3\kappa u} \right)\;.
\label{dyn}
\end{eqnarray}
The point is that this has a universal contribution coming from the
surface gravity, plus complicated sub-dominant terms that depend on the
details of the collapse.  It is important to note, however, that the
corresponding additional contributions to the RSET are finite, in
contrast to that associated with the first term. Indeed, for
small values of $x$,
\begin{equation}
u \approx t-\frac{1}{\kappa}\,\ln x +\mbox{const}~,
~~~~~~~{\rm or}~~~~~~~
\e^{-\kappa u}\propto x\;\e^{-\kappa t}~,
\end{equation}
and so the second term in the right-hand side of equation
(\ref{dyn}) is $\mathcal{O}(x^2)$, and by equation (\ref{xx}) gives
an $\mathcal{O}(1)$ contribution to $T_{xx}$ that does not depend on
$x$, but depends on time as $\e^{-2\kappa t}$. In addition, from a
comparison of equations (\ref{t0-tH})--(\ref{expp}) we see that
\begin{equation}
{A_2\over A_1} \propto {1\over\lambda}\;,\qquad {A_3\over A_1}\propto {1\over \lambda^2}\;,
\end{equation}
so the leading sub-dominant term in the RSET is inversely
proportional to the square of the speed with which the surface of
the star crosses its gravitational radius. In particular, at horizon
crossing, that is at $t=t_{\rm H}$, the value of the RSET can be as
large as one wants provided one makes $\lambda$ very small. This
would correspond to a very slow collapse in the proximity of the
trapping horizon formation.  Thus, there is a concrete possibility
that (energy condition violating) quantum contributions to the RSET
could lead to significant deviations
from classical collapse when a trapping horizon is just about to
form.

\vspace{0.2cm}
\noindent
{\bf Calculation assuming asymptotic horizon formation}
\vspace{0.2cm}

Another interesting case one may want to consider is one in which
the horizon is never formed at finite time, but just approached
asymptotically as time runs to infinity. In particular, in
reference~\cite{quasi-particle-prl} it was shown that collapses
characterized by an exponential approach to the horizon,
\begin{equation}
r(t) = 2M + B \e^{-\kappa_{\rm D} t}\;,
\end{equation}
lead to a function $p(u)$ of the form
\begin{eqnarray}%
p(u)=U_{\rm H} - A_1 \e^{-\kappa_{\rm eff} u}\;,
\end{eqnarray}%
where $\kappa_{\rm eff}$ is half the harmonic mean between $\kappa$
and the rapidity of the exponential approach $\kappa_{\rm D}$,
\begin{equation}
\kappa_\mathrm{eff} = {\kappa \; \kappa_{\rm D}\over \kappa + \kappa_{\rm D}}\;,
\end{equation}
so that one always has $\kappa_{\rm eff}<\kappa$. In this case, the
calculation of the dynamical part of the RSET leads to exactly the
same result as when using expression (\ref{standard-p}), modulo
the substitution of  $\kappa$ by $\kappa_{\rm eff}$. However, the
non-dynamical part of the RSET remains unchanged. This implies that
now, at leading order
\begin{eqnarray}%
\hbox{RSET}(x \approx 0)
\approx {1 \over \kappa^2 x^2}\left(\kappa_{\rm eff}^2-\kappa^2\right)
= -\frac{\kappa\left(2\,\kappa_{\rm D}+\kappa\right)}{\left(\kappa_{\rm D}+\kappa\right)^2\,x^2}\;,
\end{eqnarray}%
which obviously diverges in the limit $x \to 0$.  We stress that
this result does not contradict the Fulling--Sweeny--Wald
theorem~\cite{FSW}, as the calculation applies only outside the
surface of the star (\ie, for $x\geq\xi(t)$), and so the divergence
appears only at the boundary of spacetime.  Nevertheless,
particularizing to $x=\xi(t)$, this again indicates that there is a
concrete possibility that energy condition violating quantum
contributions to the RSET could lead to
significant deviations from classical collapse when a trapping
horizon is on the verge of being formed.

\vspace{0.2cm}
\noindent
{\bf Summary of the calculations}
\vspace{0.2cm}

The previous two calculations point out that the building up of a
significant RSET during the collapse, and therefore the appearance of
quantum modifications to the collapse, depend in turn on the
characteristics of the very collapse. If for whatever reason the
collapse starts to significantly deviate from free fall, then, the
building up of an energy-violating RSET could take over and further slow down 
the collapse. Surprisingly, the behaviour of the RSET is such
that it allows two rather different scenarios to be conceptually
self-consistent. One is the standard scenario: One assumes almost
free-fall collapse and the RSET results in negligible values thus
confirming the initial hypothesis. The other is the one presented
here: One assumes significant deviations from free-fall and the RSET
results in large values qualitatively consistent with the starting
hypothesis.


\begin{theacknowledgments}
Financial support for CB was provided by the Spanish MEC
through the project FIS2005-05736-C03-01 and by the Junta de Andaluc\'ia 
Excelence Project FQM2288.
\end{theacknowledgments}


\end{document}